\RequirePackage{fix-cm}
\documentclass[smallextended]{svjour3}       
\smartqed  

\usepackage[normalem]{ulem} 
\usepackage{graphicx}
\usepackage{multirow}
\usepackage{todonotes}
\usepackage{tcolorbox}
\usepackage{hyperref}

\newcommand{\interviewquote}[2]{\begin{quote}
\footnotesize{\emph{``#1'' }} --- \footnotesize{#2}
\end{quote} }

\begin{document}
\title{We Do Not Understand What It Says -- Studying Student Perceptions of Software Modelling}
\titlerunning{Studying Student Perceptions of Software Modelling}
\author{Shalini Chakraborty \and Grischa Liebel}
\date{January 2023}

\institute{S. Chakraborty \at
              Reykjavik University\\ Menntavegur 1, 102 Reykjav\'{i}k, Iceland \\ ORCID: 0000-0002-9466-3766 \\
              \email{shalini19@ru.is}
\and 
G. Liebel \at
              Reykjavik University\\ Menntavegur 1, 102 Reykjav\'{i}k, Iceland \\ ORCID: 0000-0002-3884-815X \\
              \email{grischal@ru.is}      
}

\date{Received: date / Accepted: date}

\maketitle
\begin{abstract}
\textbf{Background: }Despite the potential benefits of software modelling, developers have shown a considerable reluctance towards its application. 
There is substantial existing research studying industrial use and technical challenges of modelling.
%
However, there is a lack of detailed empirical work investigating how students perceive modelling. 
\textbf{Aim:} We investigate the perceptions of students towards modelling in a university environment.
\textbf{Method:} We conducted a multiple case study with 5 cases (5 courses from 3 universities) and two units of analysis (student and instructor).
We collected data through 21 semi-structured interviews, which we analysed using in-vivo coding and thematic analysis.
\textbf{Results:} Students see some benefits of modelling, e.g., using models for planning and communicating within the group.
However, several factors negatively influence their understanding of modelling, e.g., assignments with unclear expectations, irregular and insufficient feedback on their models, and lack of experience with the problem domains.
\textbf{Conclusions: }Our findings help in understanding better why students struggle with software modelling, and might be reluctant to adopt it later on.
This could help to improve education and training in software modelling, both at university and in industry.
Specifically, we recommend that educators try to provide feedback beyond syntactical issues, and to consider using problem domains that students are knowledgeable about. 
\end{abstract}

\section{Introduction}
\label{sec:Intro}

Software modelling has the potential to improve on several factors in software and systems engineering, such as productivity \cite{agner12} or cost \cite{kirstan10}.
While these benefits seem to be appreciated in certain areas of software and systems engineering \cite{liebel18survey}, software modelling is not widely adopted in software and systems engineering as a whole \cite{gorschek2014use}.

The reasons for a lack of adoption and use have been studied in depth, revealing issues such as poor quality code generation, lack of tool support, and lack of guidance or training \cite{Forward2010perceptions,Whittle2013Industrials,Whittle2014StateofPractice,Mohagheghi13Help,liebel18sosym,liebel18survey}.
Additionally, it has been put forward that a perceived lack of usefulness or too high effort prevents engineers from using models \cite{gorschek2014use,torchiano13}.

In the educational domain, substantial work exists on how to teach modelling in a university curriculum, e.g., \cite{Westphal2019Teaching,Loli2018Teaching,kolovos2016towards,paige2014bad,whittle2011mismatches,Stikkolorum2015LogViz,Liebel2016Impact}.
Existing work typically comes in the form of suggested course designs, e.g., \cite{schmidt2014teaching,Westphal2019Teaching}, experience reports that discuss challenges or good practices, especially tool-related,, e.g., \cite{Loli2018Teaching,kolovos2016towards,paige2014bad,whittle2011mismatches}, or papers that report quantitative studies of student opinions, e.g., \cite{Stikkolorum2015LogViz,Liebel2016Impact}. 
This body of work forms a rich source of experience and inspiration for teaching modelling.

However, there is a lack of in-depth studies that aim to understand how students perceive modelling and why this is the case.
Experience reports commonly suffer from various biases, as they rely on individual experiences and lack rigorous methods of data collection and analysis.
Similarly, existing quantitative studies aim to provide a broad picture and can, therefore, not provide detailed explanations \cite{stol2018abc}.
Finally, experience reports by researchers that have a particular focus on software and systems modelling risk being unrepresentative of instructors without such a focus.
Since students will after graduation become professionals, it is important to understand their perceptions.
Understanding existing challenges they face, but also perceived benefits of using models can help improve modelling education and facilitate an increased uptake in industry.

To address this gap, this paper aims to investigate students' perceptions of modelling, both in terms of benefits of and challenges with modelling.
We pose the following two research questions (RQs) to address our aim:  
\begin{itemize}
    \item RQ1: What are students' perceptions about modelling? 
    \item RQ2: What are the main challenges students face while modelling?
\end{itemize}


%
%
%

%
To answer these questions, we conducted a qualitative case study, interviewing a total of 21 subjects from 5 university courses that cover modelling.
We interviewed both students and instructors, and analysed course assignments.


Our findings show that several of the perceived benefits and challenges align well with those reported in existing literature.
Students find it beneficial to use models as a means of communication and handling complexity.
They report difficulties such as obtaining good and fast feedback on a model.
We further find that the lack of technical skills and domain knowledge in the student population prevents them from connecting their models and judging their quality to any known domain.
This finding gives a better explanation of why certain problem domains might be problematic in any given context.
Further, it highlights the importance to carefully tailor modelling courses to the experience of the audience, and consider the role of modelling courses in the curriculum.

The rest of the paper is structured as follows. In  Section~\ref{sec:Relatedwork} we discuss the related work on modelling in education and industry, and the importance of taking students' perception into account. In Section~\ref{sec:Method}, we present the method we applied for our case study, followed by the results in Section~\ref{sec:Result} and a discussion thereof in Section~\ref{sec:Discussion}. Finally we conclude the paper in Section~\ref{sec:Conclusion} with a summary and plans for future work. 
\section{Related Work}
\label{sec:Relatedwork}
In the following, we describe related work that investigates the use of models in industry, and work on modelling education. Finally, we describe work that aims to connect industry and education.
\subsection{Models in Industry}
There is substantial work on the adoption and the challenges of software modelling in industry, e.g., \cite{Hutchinson2011Empirical,Hutchinson2011MDEpractice,HUTCHINSON2014social,Whittle2014StateofPractice,liebel18survey,liebel18sosym,torchiano13,Mohagheghi08Proof,Mohagheghi13Help,baker2005model}.

In an early study at Motorola, Baker et al.~\cite{baker2005model} find that models increase productivity and reduce defects.
However, they also find that modelling tools are insufficient and lack interoperability, and that MBE does not scale sufficiently.

Mohagheghi et al.~\cite{Mohagheghi08Proof,Mohagheghi13Help} highlight the potential of using models for simulation and testing, while also mentioning tool issues and model complexity as challenges of adopting MBE.

Hutchinson et al.~\cite{Hutchinson2011Empirical,Hutchinson2011MDEpractice,HUTCHINSON2014social,Whittle2014StateofPractice} conduct several studies assessing the state of practice of MBE in industry.
The authors confirm that modelling tools and model complexity are problematic, but also highlight organisational factors.
Among others, they find that a lack of training hinders the adoption of modelling.

In a survey in the Italian software developing industry, Torchiano et al.~\cite{torchiano13} find that developers mainly use models for informal sketches and communication.
They further report that inexperience among developers limits the use of models. 

In a survey among systems engineering professionals, we find that models provide substantial benefits, e.g., in terms of increases in productivity \cite{liebel18survey}.
However, several challenges in technical and non-technical areas remain, e.g., lack of training and guidance, tool shortcomings, and a lack of tool interoperability.
Following up with an in-depth qualitative study at two automotive companies, we report that models are used primarily for communication and to handle complexity.
Stakeholders further prefer sketches and informal models \cite{liebel18sosym}.
System models often use in other domains, such as healthcare, for communication and to provide better organisational support. In \cite{kopach2007applying}, authors discuss the application of event-based models such as Petri nets to capture the various operations running in a  hospital. Although found helpful, the authors also highlight certain challenges like needing more understanding in adopting the models from healthcare associations. 
Considering UML specifically, the healthcare domain has multiple evidence of using UML models to design the domains \cite{walderhaug2008experiences,raistrick2004applying} and human-centric tasks \cite{bernonville2005contribution}. However, in the latter, the authors mentioned a need for more explicit representation from the models. In the healthcare domains where the cognitive behaviour of the actors is highly involved in the process, software models that are used for designing those processes should provide the ability to include the behaviours. 

\subsection{Models in Education}
Research on models in education exists primarily in three forms: (i) solution proposals on how to teach modelling, including tools tailored to an educational context, (ii) experience reports on modelling education, and (iii) surveys among students.

An example of the first category is the practical approach to teaching model-driven software development proposed by Schmidt et al.~\cite{schmidt2014teaching}.
In the proposed course, students develop a code generator using standard software development tools.
Similarly, Westphal~\cite{Westphal2019Teaching} describes the design of an SE course that focuses heavily on modelling.
In \cite{gilson2018teaching}, authors proposed a course where they used students both as language designers and its users to evaluate the usability of software language engineering (SLE). 
Gonnord et al.~\cite{gonnord2018practicing} try a reverse approach, and with the students, they journey from low-level C code to designing modelling language workbench. 
In all cases, evaluation of the proposed course design relies on student feedback.

Numerous experience reports exist related to modelling education, e.g., \cite{akayama2013tool,paige2014bad,kolovos2016towards}.
Akayama et al.~\cite{akayama2013tool} share their experience and opinion on tool use in software modelling education.
The paper describes different approaches taken by the individual authors and provides a discussion of factors such as modelling tools vs pen and paper, the conflict between the concepts of design and programming, and how to measure the quality of models.
Paige et al.~\cite{paige2014bad} discuss what they consider to be bad practices of teaching modelling.
They name bad practices such as covering a too broad range of modelling-related topics, and focusing on syntax instead of semantics.
Similarly, Kolovos and Cabot~\cite{kolovos2016towards} present a corpus of use cases for courses teaching MBE.
The authors state that modelling courses regularly suffer from the use of uninteresting or irrelevant examples, and therefore propose a list of use cases suited to teach modelling.
Daniel L. Moody~\cite{Moody2010physics} mentions a need for more effort in designing visual syntax for notations. Therefore, propose a tutorial that defines a " set of principles for designing cognitively effective visual notations: ones that are optimised for human communication and problem solving". 
Ciccozzi et al.~\cite{Ciccozzi2018How} present a survey among educators on how modelling is taught.
The authors find that educators see the focus on tools critical, as it might prevent students from understanding the core principles of modelling.

Several empirical studies exist on modelling education, e.g., \cite{Reuter2020UMLproblem,Stikkolorum2015LogViz,Agner2019Student,Liebel2017CaseTool,hammouda2014case}.
Reuter et al.~\cite{Reuter2020UMLproblem} study students' problems with UML diagrams over two modelling courses.
As a result, the authors present a catalogue of problems with UML diagrams.
In a similar direction, Stikkolorum et al.~\cite{Stikkolorum2015LogViz} study student problems and modelling strategies in UML Class diagrams by presenting students with a specialised UML editor that incorporates feedback mechanisms.
The results reveal four distinct strategies, and a number of problems such as choosing the right syntax elements.
Agner et al.~\cite{Agner2019Student} conduct a survey on modelling tool use among 117 students. The authors report issues such as a lack of feedback, difficulties in drawing the diagrams and tool complexity.
Hammouda et al.~\cite{hammouda2014case} compare the use of modelling CASE tools to pen and paper use.
Using a survey, they evaluate students' perceptions of the differences, finding no clear advantage for either approach.
In the context of modelling tools in education, the tool Umple needs to be highlighted \cite{garzon2015umple}.
Umple is a tool that allows to create UML models in a textual concrete syntax close to object-oriented languages like Java.
Furthermore, code in many different general-purpose programmming languages can generated directly from Umple models, thus allowing for direct feedback from models.
Surveys with students have shown positive results on the use of Umple \cite{lethbridge2011teaching,Liebel2017CaseTool}.
Umple is successfully used by several instructors teaching modelling.
However, to our knowledge, it is currently not widely used on a global scale.
Finally, we conducted several case studies on tool use in modelling education \cite{Liebel2017CaseTool,Liebel2016Impact}.
We find that students can use industrial modelling tools successfully, but require substantial coaching in how to use the tools, with a dedicated tool champion present in the course.
We further find that the tools' inability to provide adequate feedback impacts the tool acceptance. 
In an attempt to connect industry practice to education, Whittle and Hutchinson~\cite{whittle2011mismatches} present essential differences between industry practice and education concerning software modelling.
The authors find that modelling education is more UML-centric, whereas industry is placing more emphasis on abstract models. 
Furthermore, in industry it is more common to use a bottom-up approach to adoption, whereas modelling is typically taught in a top-down fashion.
\section{Research Method}
\label{sec:Method}
Our goal is to investigate students' perceptions, especially their challenges in learning and applying software modelling in a university environment.
To do so, we conducted a multiple-case study.
Runeson at al. \cite{bookRuneson2012} define a case study as an ``empirical investigation of a software engineering phenomenon within its real-life context'' where evidence can be drawn from multiple sources, more importantly from real-life experiments involving human participants \cite{WOHLIN2021Casestudy}.
The phenomenon we study is the experience of students learning software modelling at university. 
In total, we interviewed 16 students from 3 different universities and 5 different courses. Additionally, we interviewed instructors of the 5 courses we considered.
Three courses were at bachelor level and two were from masters level. %
In addition to the interviews, we consulted assignment details for the courses.

Whether students can or should be used as study subjects is a long-debated question in SE \cite{Runeson2003graduate,Salman2015Are,sjoberg2002conducting,Falessi2018ExpertsonStudents}.
In particular, students can be valid study subjects under certain conditions, e.g., if novice software engineers are studied \cite{Falessi2018ExpertsonStudents}.
In our case, we aim to study how students perceive modelling, in contrast to existing studies that investigate industrial cases.
Therefore, the choice of students as study subjects is valid.

The study setup is described in detail in the following sub-sections.


%
\subsection{Case Study Design}
\label{sec:Study_Design}
\subsubsection{Pilot Study} Before starting the actual study, we conducted a pilot study with three interviewees, two doctoral students and one bachelor student. Of the two doctoral students, one has prior knowledge and experience of software modelling. The other student has no experience in software modelling. 
We selected the combination because we wanted to check how our questions would be received by different persons with various software modelling background, as we are dealing with students from different countries, courses and backgrounds. Then we conducted another interview with a bachelor student who had taken a modelling course at our university. Our focus was to check the interview guide. Also, the pilot study helped us to estimate the overall interview time.   

\subsubsection{Cases and Recruitment}
We designed a multiple \emph{embedded} case study \cite{bookRuneson2012} with five cases, each case is represented by a course-university pair. Table~\ref{tab:case_summary} shows the cases and the interviewees per case (using anonymous identifiers).
Initially, we contacted the instructors of the courses based on our own contacts, and once they agreed, we advertised the study to students of that course.
C1 was selected through convenience sampling, as it is a course at our home university.
It is a typical ``UML course'' in the sense that basic UML diagrams are introduced. The course starts with user interface methods, later students learn about software modelling, including the UML use case, class, sequence, and activity diagrams. Assignments related to these diagrams are given. The course has two instructors, where I2 was responsible for the modelling part. However, both I1 and I2 were in charge of the exams. Therefore, we interviewed both instructors. 
Neither of the two instructors is active in modelling research.
C2 was selected as a comparative case, as it was a basic UML course taught by the same instructor I2 as C1, at the same university. 
C2 concentrates mainly on software modelling, with more focus on UML in the assignments.
Similarly, the difficulty level is higher.
C1 and C2 are both bachelor courses. However, C2 is located in the software engineering program and  C1 in computer science. 
C3 to C5 were selected as revelatory cases, as they exhibit one or several differences to C1 and C2.
One of the main differences is the change of university/country. C2 and C3 are similar in terms of the course syllabus. Both courses teach system design and UML from scratch.
C4 and C5 are courses in the same university and are master courses. While, in C4, UML is taught from scratch, C5 focuses on using UML diagrams for architecture and testing. 
Apart from C1, all four courses include implementation tasks.

Participation in the interviewees was voluntary and instructors were not aware of who participated in the interviews. 
%
We asked interested students to contact the researchers via mail, and once students initiated the contact, we sent them an interview meeting invitation and a consent form. All interviews were online. To appreciate students' participation, we offered them movie vouchers or did a donation to charity on their behalf. 
\begin{table*}[ht]
\caption{Case Summary}
\label{tab:case_summary}
\centering
\begin{tabular}{p{0.04\linewidth}p{0.08\linewidth}p{0.5\linewidth}p{0.18\linewidth}p{0.08\linewidth}}
\hline
Code & Level & Course Description & Students & Instructors\\
\hline \hline
C1 & Bachelor & Software analysis and prototype design using user interviews & C1\_(S1, S2, S3, S4, S5) & I1, I2 \\
\hline
C2 & Bachelor & Software requirement analysis and design & C2\_(S1, S2, S3) & I2 \\
\hline
C3 & Bachelor & Software requirement specification and design & C3\_(S1, S2) & I3\\
\hline
C4 & Master & System specification, design and testing & C4\_(S1, S2, S3) & I4\\
\hline
C5 & Master & Software architecture design and quality analysis & C5\_(S1, S2, S3) & I5\\
\hline
\end{tabular}
\end{table*}

\subsection{Data Collection}
\label{sec:Datacollection}
Based on the pilot study, we decided to conduct interviews for approximately 40 minutes.
The student interviews were divided into two set of questions: \textbf{introductory question} and \textbf{model questions}. In the \textbf{introductory questions}, we added questions about a student's background, previous work experience with software models. The \textbf{model questions} includes students' personal experiences during courses, challenges in learning modelling, completing assignments and overall perception of software modelling.
The instructor questionnaire includes details about the course structure, assignments, feedback from students and the instructor's challenges with the course.
All interviews conducted remotely through Zoom or Microsoft Teams, and their build-in recording features.
Before recording, we asked permission from each interviewee and received their signed consent form. 
Interviewees could additionally give their consent that we would publish their anonymised interview transcript.
13 interviewees consented to publishing.
The interview guides and the transcripts can be found on Zenodo~\cite{chakraborty_shalini_2022_6913780}.

\subsection{Data Analysis}
\label{sec:Analysis}
\begin{figure*}
    \centering
     \includegraphics[width=\linewidth]{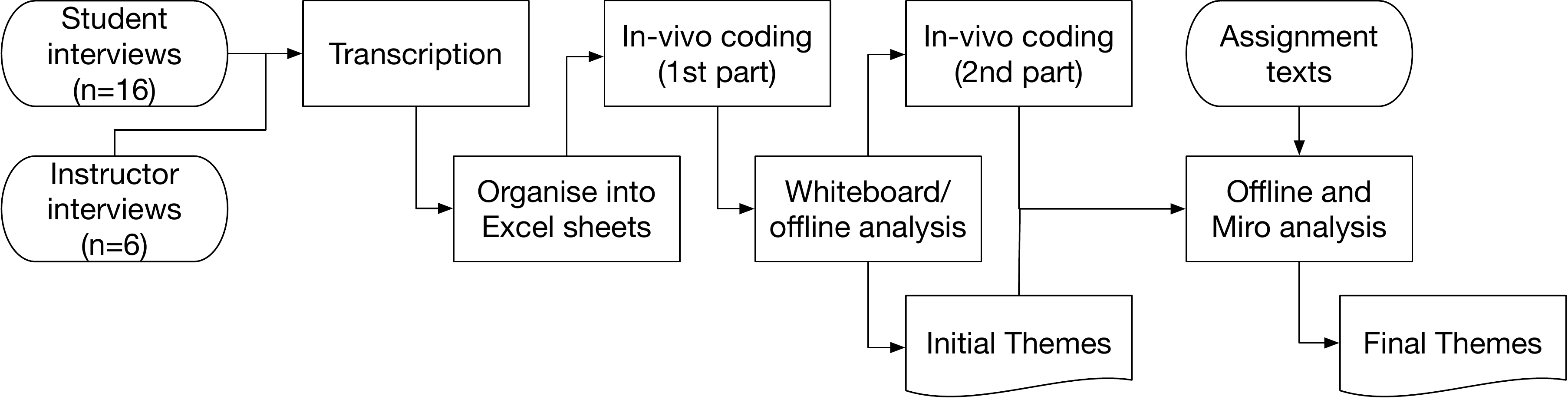}
	\caption{Data Analysis Process. Rounded rectangles denote data sources, regular rectangles denote activities, rectangles with wavy bottom line denote artefacts, and arrows denote information flow.}
	\label{fig:data_analysis}
\end{figure*}

The data analysis process is depicted in Figure~\ref{fig:data_analysis}.
After conducting the interviews (steps 1a and 1b in Figure~\ref{fig:data_analysis}), we transcribed the interviews using transcription services\footnote{We used Konch\footnote{https://www.konch.ai/} and Go Transcript\footnote{https://gotranscript.com/}.}.
In case of automated transcriptions, we post-processed the transcripts to improve their quality.
The two researchers then applied in-vivo coding separately on each interviewee transcript.
We used in-vivo coding \cite{Saldana2015codingmanual}, as it helps to highlight participants' opinions by using the actual spoken words.
In our case, interviewees were from different countries and cultural backgrounds, and thus used varying vocabulary to describe their personal experiences.
After conducting the first five interviews (C1\_S1, C1\_S2, C1\_S3, C1\_S4 and C2\_S1), we coded and then jointly discussed the resulting codes, grouping them into initially 10 themes.
These themes related to benefits and challenges of modelling, as well as general categories refering to course feedback and concerns.
We refined the themes as analysis progressed and more interviews were added. 

Once we finalised the themes based on student data, we coded the instructor data and compared it with the existing themes.
We used the instructor interviewees for three purposes, namely (a) to understand the course context, (b) to better understand the student perceptions and to check whether the instructors had the same views and (c) to obtain ideas for potential best practices.
For sorting quotes and themes, we used the online whiteboard tool Miro\footnote{https://miro.com/}.

As a second data source for our analysis, we used course assignments to cross-check themes and quotes that related to, e.g., assignments, work load, or grading.
We contacted the instructors after their respective interviews for assignment details, which included the assignment structure, time, instructions, and point distribution.
Unfortunately, only three instructors (C1/I1, C3/I3 and C4/I4) responded. However, these three courses consist of the majority of the students. For the other two courses, we used the instructor interview data, and tried to gather as much information about course timing, assignments, and grading (see questions 3 and 4, in Appendix~\ref{app:Interview_Guide}2).  

We checked the assignments for the problem motivation, the actual tasks, the required problem domain knowledge, clarity of instructions and the total time dedicated to modelling tasks.
We then compared these aspects to statements made during the interviews.

The total time of data collection and analysis was approximately one year.
The resulting themes related to modelling benefits are listed in Table~\ref{tab:student_themes_benefits}, with example quote for each theme.
Similarly, Table~\ref{tab:student_themes_challenges} lists the themes related to modelling challenges.


%

%
%
\begin{table*}[ht]
   \caption{Final themes related to modelling benefits, with a short description and a sample quote associated with each theme.}
    \label{tab:student_themes_benefits}
\centering
\begin{tabular}{p{0.15\linewidth}p{0.34\linewidth}p{0.4\linewidth}}
   \hline
    Theme & Description & Example statement \\
    \hline \hline
    B: Same Page & Benefits of modelling in communication and coordination within teams. & \textit{I think it's important for documentation, as well as getting everyone on the same page} \\
    \hline
    B: Better Planning & Modelling helps in planning ahead, for understanding the requirements, potential design alternatives, but also to plan implementation. & \textit{But the thing I liked this is when you feel like you have a plan and you know what are you gonna do} \\
    \hline
    B: Better Understanding of Code & Benefits of modelling to help understanding a code base without reading the code in detail. & \textit{I really like the idea of seeing it as a language to talk about code, because I will never read someone's code if they ask for feedback-} \\
    \hline
   B: Maintenance/ Documentation & Benefits of modelling as a way to help maintaining a system and to document what it does. & \textit{Usually, it's other people who will maintain software, so they need to read what you make if they need to understand.} \\
    \hline
    B: Personal Preferences & Benefits related to individual preferences. & \textit{Yeah, yeah we used that because it was um..prettier, than this is the whole perfectionism thing with um, it was prettier than coding}\\
    \hline
    B: Doubtful Benefits& Doubts regarding any benefits of modelling.& \textit{[..] ]you can't be mastering everything and [I am] more into implementation rather than modelling things}\\
        \hline
       \end{tabular}
\end{table*}
    
\begin{table*}[ht]
   \caption{Final themes related to modelling challenges, with a short description and a sample quote associated with each theme.}
    \label{tab:student_themes_challenges}
\centering
\begin{tabular}{p{0.15\linewidth}p{0.34\linewidth}p{0.4\linewidth}}
   \hline
    Theme & Description & Example statement \\
    \hline \hline
   C: Unclear Expectations & Confusion on what the purpose of modelling is, often as expectations are not clearly communicated. & \textit{At the end it's like 'Oops, maybe I should have designed', but then I wouldn't know how to design it.} \\
    \hline
    C: Irregular and Unclear Feedback & Feedback is provided too seldom, is not clear, or is restricted to formalities, such as the diagram syntax. & \textit{Writing for hours and then a teacher be like no this is wrong}\\
    \hline
    C: Lack of Expertise in the Problem Domain & Students are provided with assignments in a domain they are unfamiliar with. Therefore, they struggle to relate their learning to something known. & \textit{we didn't have much experience in programming and designing and modeling the app, I think the biggest challenge was in the first week that we were just flowing with our ideas how the app should look like without actually knowing how the end picture of the class diagrams should look like}\\
    \hline
    C: Time and Repetition & Modelling consumes time, and requires repetition to master. These two factors are often in conflict in university courses, and students struggle to see the value of repeatedly improving their models..& \textit{Later on, we had to fix it at least 10 times. After writing the app, we had to go back to our diagrams and redo them how our final vision was} \\
    \hline
    C: Notation & Struggles with the complexity of the UML notation. & \textit{One thing that comes up is with the state diagram and activity diagram, which one is supposed to do what}\\
    \hline
    C: Tooling & Modelling tools can cause numerous difficulties, such as poor usability. & \textit{I hate PlantUML. I can't read the diagram that comes out of it. It's all over the place} \\
    \hline
    C: Lack of Cooperation & Challenges due to difficulties in collaborating in teams, and due to lack of professionalism in students' attitudes. & \textit{[..] one or two would make the model and the rest maybe don't understand. Somehow, I don't know how to solve this problem.}  \\
    \hline
   \end{tabular}
\end{table*}

\subsection{Validity threats}
\label{sec:Validity}
We conducted an exploratory \cite{bookRuneson2012} case study, where we primarily collected data through interviews. For the analysis, we leaned towards interpretivism. An interpretivist approach means that \emph{``humans construct knowledge as they interpret their experiences of and in the world; rejecting the objectivist notion that knowledge is simply there to be identified and collected''} \cite{hiller2016epistemological,pascale2010cartographies}. In our work, adopting to interpretivism is suitable as we seek answers for our RQs through the interviewees' perspective, based on their knowledge of the addressed subject and cultural background. The understanding of our knowledge is therefore relative to the person and their personal experience. 

Following the work of Petersen et al.~\cite{petersen2013worldviews}, we present the validity threats of our work and the measures we have taken to mitigate them in the following. 

\subsubsection{Transferability}
Transferability describes to what extent results from the study can be transferred to cases that resemble the case under study \cite{petersen2013worldviews}.
Cultural differences, the influence of teachers and their teaching practices, and the course subjects limit transferability in our study.
We aim to ensure a substantial level of transferability by basing our analysis on five cases.
Nevertheless, all five cases teach modelling for analysis tasks on a high level of abstraction, i.e., for requirements, architecture and design purposes.
Specifically, none of the courses covers model transformation or other tasks that require formal models.
Furthermore, all five courses are located in Northern European Universities, thus potentially limiting the applicability to other countries.
Finally, perceptions of modelling are tightly connected to perceptions of the course.
Specifically, the quality of teaching and the instructor's expertise on the topic could positively or negatively affect the studied perceptions.
To avoid this, we selected both cases where the instructors do and do not have a research background in software modelling.
Pedagogical quality is harder to assess.
However, we did not get the impression from students that individual courses had a low level of pedagogical quality.
Nevertheless, this might be a threat to transferability.

To further allow for transferability, we conducted a substantial number of interviews (21).
We reached a saturation point in our themes after the 16 student interviews, and hence stopped including further cases or interviewees.
That is, in the last batch of interviews (n=6) we analysed, no new themes emerged that we had not yet included in our analysis.

\subsubsection{Credibility}
Credibility describes to what extent findings have been distorted by the researchers \cite{petersen2013worldviews}.

We tried to avoid distortion of the reported findings by grounding the analysis in in-vivo codes directly taken from the verbatim interview transcripts.
Furthermore, we performed member checking to get feedback on the extracted themes from our interviewees.
Only few interviewees answered this call, but those confirmed the credibility of the found themes.

All interviews were recorded, and data analysis performed on the verbatim transcripts.
Additionally, we publish the anonymised transcripts of the 13 interviewees who consented to this.
This should ensure credibility of the findings.

\subsubsection{Confirmability}
Confirmability describes the extent to which conclusions made by researchers follow from the observed data \cite{petersen2013worldviews}.

To allow for confirmability, we presented the way of coding in depth.
Furthermore, we give example quotes for each theme, and publish anonymised student transcripts.
Note that this does not necessarily ensure reliability, i.e., that analyses conducted by other researchers would yield precisely the same results.
We discussed themes and codes and aimed to find consensus in our analyses.
Nevertheless, we allowed for some disagreements to account for subjective impressions or opinions, as is common in interpretivist research.
\section{Results}
\label{sec:Result}
%
In this section we report and discuss the findings of our case study. 

\subsection{Student Perception of Software Modelling (RQ1)}
\begin{figure}[ht]
    \centering
    \includegraphics[width=.75\linewidth]{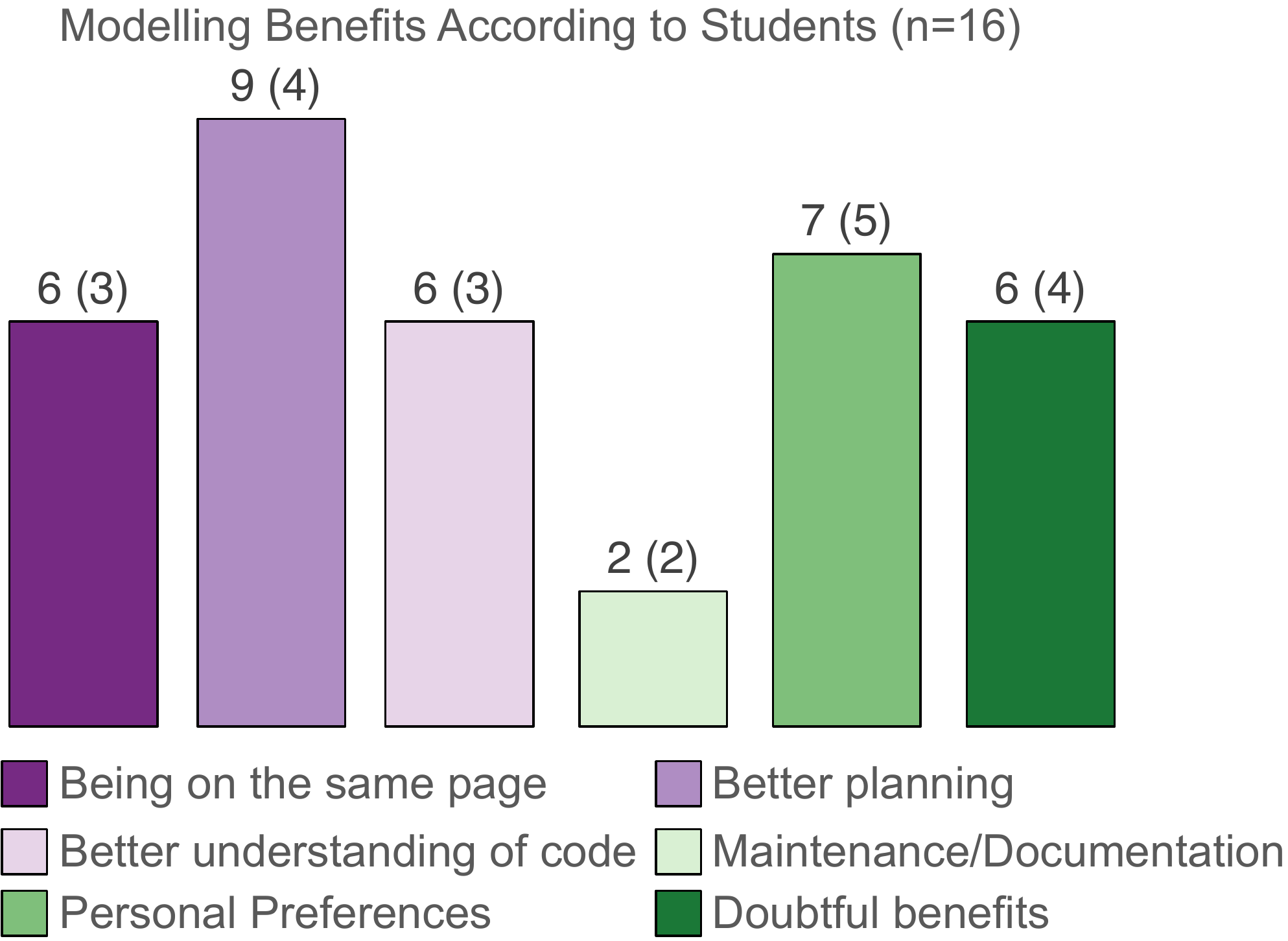}
    \caption{Benefits as supported by the student interviewees. Numbers over the bars represent the total amount of students mentioning the benefit, and the number of cases in which students mentioned the benefit in parenthesis.}
    \label{fig:benefitsStudents}
\end{figure}

Based on the interview data, we observe that students see specific benefits of modelling, primarily for areas where informal models might be sufficient, such as obtaining a system overview or conceptual understanding of the problem domain. 
However, many are skeptical as to what value modelling has for detailed system design.
In the following, we discuss the students' perceptions and provide adequate interview quotes that support the categories.
An overview of how many students mentioned the perceived benefits is depicted in Figure~\ref{fig:benefitsStudents}.

\subsubsection{Being on the Same Page}
Several students find that modelling helps their groups communicate better with each other. Students mention modelling helps them to express their ideas, share work and make decisions as a group. For example, students stated:
\interviewquote{I think it's important for documentation, as well as getting everyone on the same page}{C2\_S1}
\interviewquote{It's a lot easier to work together when you have the diagrams.}{C1\_S5}
\interviewquote{It helps every group member to know what exactly they want}{C3\_S2}
%
\subsubsection{Better Planning}
Students find modelling helpful in planning the development, i.e., designing systems.
Students appreciate that they can visually map the system and plan the development through modelling.  
\interviewquote{But the thing I liked this is when you feel like you have a plan and you know what are you gonna do}{C1\_S2}
\interviewquote{Class diagrams are really, really helpful. Before starting your project...because you can see and have the view of your classes}{C5\_S3}
\subsubsection{Better Understanding of Code}
In relation to programming, some students found models helpful to get an overview of the code on a higher level of abstraction.
That is, they stated:
\interviewquote{I really like the idea of seeing it as a language to talk about code, because I will never read someone's code if they ask for feedback}{C3\_S1}
\interviewquote{If you have a diagram for it, we would get a better understanding of the code itself.}{C2\_S3}

An aspect of this is a top-down modelling approach, where a better understanding of the system is obtained through breaking down the system in steps, as noted by some of our interviewees.
\interviewquote{If you are making a program that has more than 200 lines of code then you probably gonna have to model it.}{C1\_S1}
\interviewquote{that was a very nice experience of seeing how you start with a kind of vague idea and then start to break it down more concrete classes}{C3\_S1}

\subsubsection{Maintenance/Documentation} 
Students find modelling to be helpful to prepare for future tasks, such as maintaining and documenting the product. However, in our interview data only students with industry background expressed this benefit.  
\interviewquote{Usually, it's other people who will maintain software, so they need to read what you make if they need to understand.}{C1\_S3}
\interviewquote{Unless you don’t have the documentation, you will end up with just a total mess}{C4\_S2}
While looking at the assignment details and interview data regarding that, we realized none of the courses explicitly mention maintenance/documentation. The assignments ask for an overview of the diagrams, but there aren't any requirements for documentation. The importance of maintaining a document or how to do that still needs to be clarified for students.  

\subsubsection{Personal Preferences}
In addition to the benefits stated above, some students use models in terms of personal preference. For example:
\interviewquote{Yeah, yeah we used that because it was um..prettier, than this is the whole perfectionism thing with um, it was prettier than coding}{C1\_S2}
\interviewquote{It helps me basically get my idea out there a lot better. Basically, when I'm starting a project, I like modeling the higher level and just going deeper and deeper and deeper.}{C2\_S1}

\subsubsection{Doubtful Benefits}
Despite experiencing benefits, several students are doubtful whether the benefits of modelling outweigh the issues, and if they will apply models in the future.
We received several statements of students saying they will most likely not apply models in the future.
\interviewquote{I think it's a great experience that we can have this now. Not in the future, in our jobs}{C3\_S2}
\interviewquote{[..] ]you can't be mastering everything and [I am] more into implementation rather than modelling things}{C4\_S2}
Interestingly, both statements above were made by students with industrial software development experience.
The same students however admitted that in their experience they were tasked with implementation only, and not with high-level tasks such as system design or requirements engineering.

One of the instructors confirmed that many of their students would question the application of modelling in industry:
\interviewquote{It's the same experience that I have when people talk about this course, they're asking, 'Is anyone using it?'}{I2}
%
 


We further discuss the challenges with modelling in the next section.

%
\subsection{Modelling Challenges (RQ2)}
\label{sec: RQ2_students_challenges}
\begin{figure}[ht]
    \centering
    \includegraphics[width=.75\linewidth]{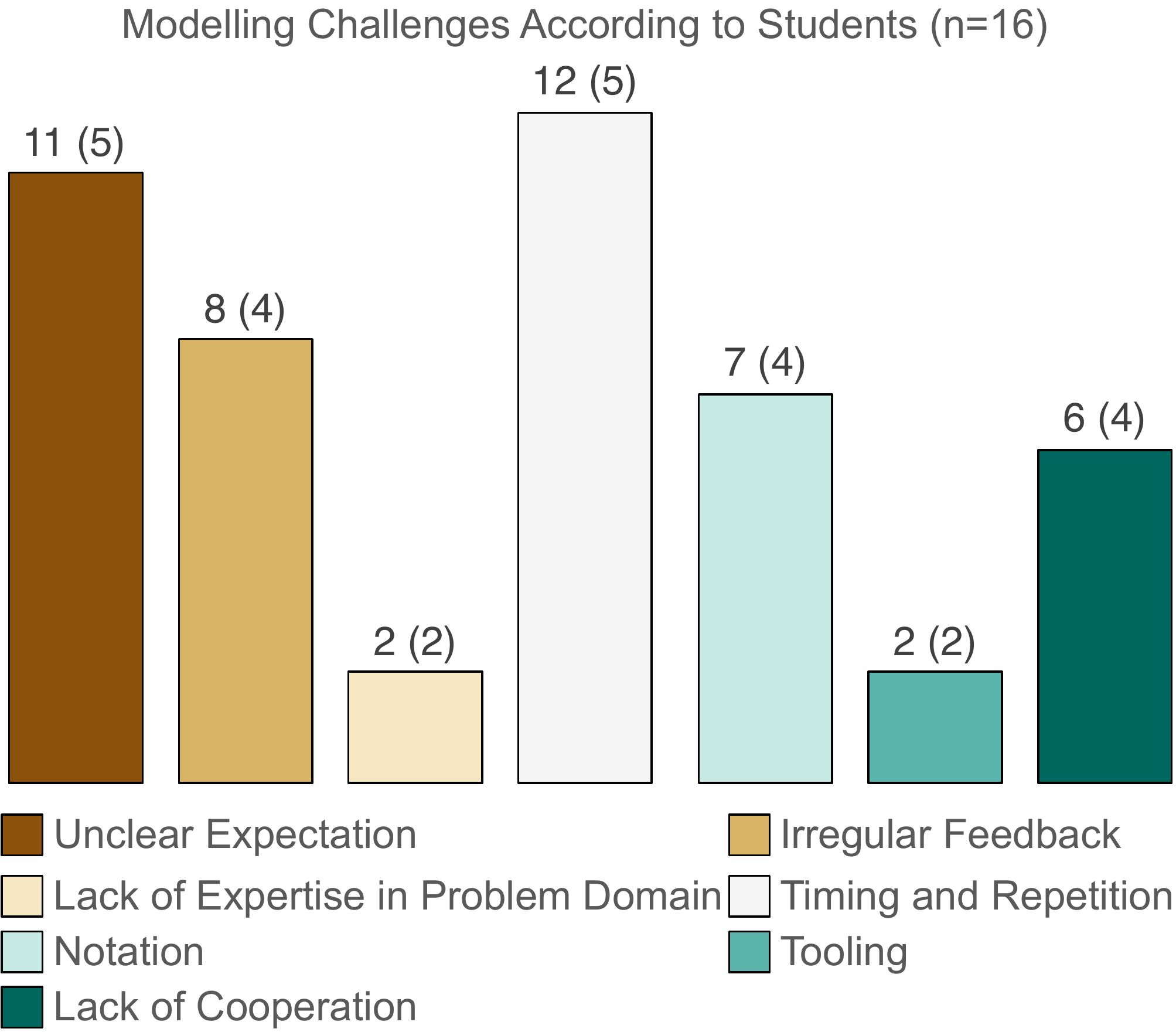}
    \caption{Student challenges as supported by the interviewees. Numbers over the bars represent the total amount of students mentioning the challenge, and the number of cases in which students mentioned the challenge in parenthesis.}
    \label{fig:student_challenge}
\end{figure}

\begin{figure}[ht]
    \centering
    \includegraphics[width=.75\linewidth]{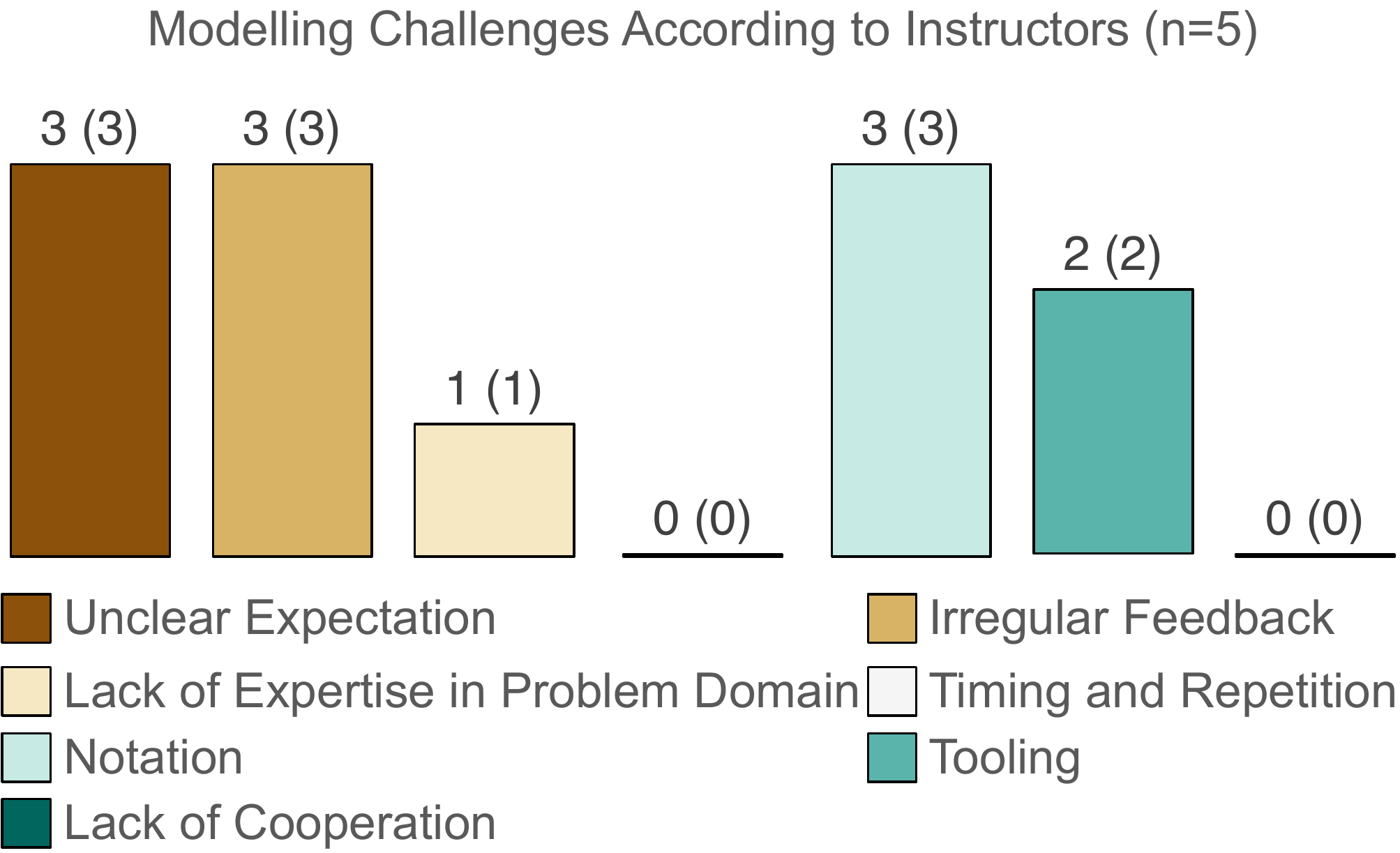}
    \caption{Student challenges as supported by the instructor interviewees. Numbers over the bars represent the total amount of instructors mentioning the challenge, and the number of cases in which instructors mentioned the challenge in parenthesis.}
    \label{fig:Instr_student_challenge}
\end{figure}

Despite benefits observed by the majority of the students, they experience several challenges related to modelling.
These relate, among others, to tooling, how to choose the right notation, what to express in the models, and how to apply it to unfamiliar domains.
%
We will discuss these challenges in the following, focusing on the student lens, and complementing their views with the instructor perspective.
Overall, we extracted 8 types of challenges.
The support by the different interviewees is depicted in  Figure~\ref{fig:student_challenge} and Figure~\ref{fig:Instr_student_challenge}.

%

%

\subsubsection{Unclear Expectations} 
In university modelling courses, students typically receive assignments to create models.
However, they often struggle to understand what is expected of them, and how to create the models.
\interviewquote{We started coding and at the end of the day, we didn't know what we did}{C4\_S3}

In particular, this is caused by a lack of knowledge in programming, software architecture, and other practical skills necessary to envision a ``good'' design:
\interviewquote{At the end it's like 'Oops, maybe I should have designed', but then I wouldn't know how to design it.
}{C3\_S1}

Additionally, assignments are often formulated in a way that is in contrast to how they are assessed, e.g., by asking students to create a \emph{prescriptive} model, which is then assessed by how closely it resembles the final code, i.e., in a \emph{descriptive} way.
Students quickly pick up this discrepancy in grading and optimise their efforts towards the grading.
That is, they take shortcuts initially, and simply update the model later to reflect their actual code.
\interviewquote{we had to, we had to basically just create a class diagram out of the code that they wrote, instead of writing a code out of the class diagram}{C1\_S1}
\subsubsection{Irregular and Unclear Feedback}
Students are dissatisfied with the type of feedback they receive.
The evaluation criteria often demand \textit{"Diagrams are well structured and provide good overview"}- assignment details from C3, \textit{"Explanation of good design suggestions"}-assignment details from C1, which is vague and do not leave space for constructive, detailed feedbacks. 
As modelling is often a qualitative task, teachers often revert to giving feedback on objective things such as the diagram syntax.
However, this is not perceived as useful feedback.
\interviewquote{So no one can say anything to us. It's good or not. Just we should follow some rules about the diagram, for example, what is solid line, what is dashed line.}{C5\_S1}

Instead, students would like more ``holding hands'' with regular feedback.
\interviewquote{Maybe because it's the first year, this is the first introduction to the subject, so people miss a lot of points, but if it had another level, teach people more, and take their hand more to do these diagrams themselves, that would be good}{C1\_S3}

In particular, since there is no automated feedback as, e.g., in programming, students require timely and regular feedback.
\interviewquote{writing for hours and then a teacher be like no this is wrong}{C1\_S1}
\interviewquote{Reply to emails more than five or six hours}{C4\_S1}

The regularity and quality of feedback was further complicated by Covid-19, as all 5 cases used remote teaching. 

\subsubsection{Lack of Expertise in the Problem Domain}
Deep knowledge of the problem domain is necessary to perform many software engineering tasks in a satisfactory manner, e.g., programming \cite{Oliveira2000Usingdomain} or program understanding \cite{rugaber2000use}.
If this domain knowledge is lacking or entirely missing, software engineering tasks cannot be performed well. Allowing students to work with a known problem increases their interest and participation.
Several students stated that, in addition to just learning modelling, they were also lacking domain knowledge and programming experience.
Therefore, they were unable to relate their models to familiar concepts.
\interviewquote{we didn't have much experience in programming and designing and modeling the app, I think the biggest challenge was in the first week that we were just flowing with our ideas how the app should look like without actually knowing how the end picture of the class diagrams should look like}{C1\_S4}

Compared to other introductory topics, such as programming, this makes modelling particularly complex: Students are expected to learn a new concept (modelling), neither having problem domain nor software design/architecture experience.
In addition, the models used in the courses we studied do not provide any kind of automated feedback, such as compiler/runtime error messages in programming.
This leads to feelings of ``just drawing something'' in students, without an anchor to connect their models to.

In a similar direction, the chosen problem domain might also affect how representative the learning is for actual systems in industry, or how suited modelling is for the given problem.
For instance, one student noted:
\interviewquote{we got to know all the basics and the whole idea and how it's supposed to look like, but not how to use it in the future on the bigger picture and the scale of the app}{C1\_S4}

\subsubsection{Time and Repetition}
Students complain that modelling takes time, and courses are tightly scheduled.
Students feel pressure due to the bulk of assignments, especially when they are not fully aware of what is expected from them. 
Given that pressure, they try to prioritise and minimise effort where possible.
However, arguably, learning how to model will require the students to make many mistakes and correct them in iterations.
Together with the lack of feedback mentioned above, students perceive these iterative improvements as a waste of time.
\interviewquote{You run it once and then probably change. It’s just a waste}{C3\_S1}
\interviewquote{we designed something, it didn't work, so we needed to change it
}{C2\_S1}
\interviewquote{Later on, we had to fix it at least 10 times. After writing the app, we had to go back to our diagrams and redo them how our final vision was}{C1\_S4}

Interestingly, these students did not perceive that iterative changes helped them understand the problem domain better, but rather saw it as a burden with unclear benefit. One of the reasons is courses contain individual assignments without a continuation or connection between themselves for some cases (C2 and C3). For each problem students are instructed to draw a specific diagram and then move on to something else. 
An exception we observed for C4 and C5 since, in C5, students are applying what they learned in C4.

\subsubsection{Notation}
Keng et al.~\cite{siau2006identifying} reported different learning challenges with UML over a decade ago. 
In our cases, we observe similar challenges.
Students find it hard to remember the syntax of different diagrams and their purpose, pointing to a lack of prior knowledge of UML. 

\interviewquote{many different types of diagrams to represent the same thing}{C1\_S1}
\interviewquote{The hardest part about drawing UML diagrams is always remembering what exactly the syntax is}{C2\_S1}

Specifically, our interviewees are most often pointing out UML state machine and activity diagrams as confusing.
\interviewquote{I have this image in my head where the state diagram is not the UI actions. I can't wrap my head around it.}{C2\_S3}
\interviewquote{One thing that comes up is with the state diagram and activity diagram, which one is supposed to do what.
}{C1\_S5}
The instructors confirm this view.
\interviewquote{here's a bit of struggle in between the concepts that they express in a diagram and the particular diagram types}{I3}
\interviewquote{they've been mixing those pretty well up [state and activity diagrams]}{I2}

One instructor offered the explanation that this relates to a lack of experience with object-oriented programming in general.
\interviewquote{they don't have a lot of experience using object-oriented programming.}{I2}
This quote reinforces our discussion above, that students are exposed to various unknown activities when learning how to model, and therefore lack knowledge they can anchor their modelling experiences to.
%
\subsubsection{Tooling}
None of the 5 cases we investigated had any requirements for using a particular modelling tool.
Correspondingly, students did not voice any tool challenges.
Students used multiple tools for their assignments, e.g., Draw.io, PlantUML, and Lucidchart. Also, some students used plain pen and paper for drawing and communicating their diagrams.
Students appreciate this choice.
\interviewquote{I think actually giving people the freedom to do what they want with the tools is really nice, because some people actually liked drawing it on an iPad or something}{C2\_S1}

In particular, some students also showed initiative at choosing different tools depending on the situation.
\interviewquote{When I did the prototype of my app, I used the Figma app on the internet. It's really good to visualize the prototype. Then we used the Lucidchart and draw.io, I think it's diagrams.net now. This is for the diagrams. Then we just used the whiteboard to draw it if we needed to because we just gathered together and we'd just throw it on the whiteboard and then we just translated it into diagrams or Lucidchart}{C1\_S4}

While offering choice in tooling clearly addresses many tool issues, it has the limitation that it only works if the course and the course assignments do not rely on specific tool features, such as code generation or simulation capabilities.
This is a limitation in our study, as all our cases introduced UML only as a means to document, plan, and to communicate, without any automated processing of models in the form of model transformation or other facilities.

\subsubsection{Lack of Cooperation}
In all 5 cases, students had to cooperate in teams.
In fact, modelling was in all cases also intended as a way to facilitate communication in the group.
However, in practice they struggled to cooperate and to appreciate the use of models for communication purposes.
\interviewquote{We are six, for example, but one or two would make the model and the rest maybe don't understand. Somehow, I don't know how to solve this problem.}{C1\_S3}
\interviewquote{Some people don't like to draw., but we're supposed to work together.}{C5\_S1}

%

%


\subsubsection{Summary}
Overall, we observe that students struggle with various aspects of models (as depicted in Figures~\ref{fig:student_challenge} and \ref{fig:Instr_student_challenge}).
Apart from classic issues such as learning an unknown notation (be it modelling or not), we find that students lack a connection to previous knowledge.
That is, due to a lack of domain, design and programming knowledge, and a lack of automated feedback from modelling tools, they cannot create an anchor.
Essentially, they produce a model that they cannot judge on any other level than notation.
One student summarised this sentiment as follows:
\interviewquote{The diagram is supposed to speak for yourself [sic]. But we didn't understand what it was saying}{RU\_G\_S1}

%
\section{Discussion}
\label{sec:Discussion}
Several of our findings that relate both to perceived benefits and challenges of models are specific to the selection of courses we studied.
That is, in all five courses, models are created for planning, documentation, design, and communication purposes.
In contrast, topics that require formal models, e.g., for code generation, formal verification, or simulation, are not included.
This certainly limits how general our findings are with respect to software modelling as a whole.
Nevertheless, we believe that a large percentage of computer science students worldwide get exposed to software modelling through similar courses, i.e., the stereotypical ``UML courses''.
As such, our findings have the potential to be transferable to many students.
Similarly, even if students take more advanced courses on software modelling, their beliefs and preconceptions about modelling can be irreversibly shaped by their first exposure to the topic.
As such, we believe our findings are particularly relevant.

In a similar direction, there is a potential threat that challenges perceived by students might not so much relate to software modelling as a course topic, but instead by the pedagogical quality of the courses.
We did not try to assess the quality of the individual courses, but at least we cover a variety of expertise in modelling on the lecturers side, ranging from instructors that do not work on models in research, to instructors that publish in software modelling venues and have or have had a particular focus on the topic.
Therefore, we do not believe that the students' experiences can simply be related to issues in teaching.

Modelling tools have been reported consistently as a concern in studies on software modelling, both related to industry and education.
In contrast, we did not find much reported challenges with tooling.
To some extent, we believe the reason for this absence of tooling challenges is the focus on ``drawing'' diagrams rather than formal modelling, and the freedom of choosing a modelling tool of the students' choice.
None of the cases used tools that are specifically designed for education, such as the Umple modelling tool \cite{garzon2015umple}.
Since surveys with students have shown positive results \cite{lethbridge2011teaching}, we believe studying such a course in depth using qualitative methods could yield further insights that are complementary to ours.
However, we also believe that the use of software modelling tools tailored towards education is currently not representative.

While this study focused purely on the educational context, it is interesting to discuss potential similarities and differences to industrial practice.

First, we observe that models are appreciated for communication purposes and to handle system complexity in our cases, something that is also reported in industry \cite{gorschek2014use,liebel18survey,storrle2017conceptual}.
Other benefits reported in industry, such as simulation or verification capabilities do not apply in our cases, since all five courses used modelling only on an informal level to express models for planning, communication, and coordination.
Similarly, it is hard to reason about improvements reported from industry, e.g., in terms of productivity \cite{baker2005model,Mohagheghi08Proof}.

Tooling issues are a frequent topic in industry, e.g., reported in \cite{Whittle2013Industrials,Whittle2014StateofPractice,Hutchinson2011Empirical,Hutchinson2011MDEpractice,Whittle2013Industrials,liebel18survey,liebel18sosym}.
In contrast, none of our interviewees raised tool issues.
One explanation for this observation is clearly that none of our courses mandated a CASE tool for modelling, but instead left that choice to the students.
Similarly, while models were often assessed for semantic correctness, the specified purposes of the models did not require syntactically or semantically correct models.
Together with a lack of requirements for interoperability between tools, this removes most of the issues contemporary modelling CASE tools have.
Nevertheless, it is positive to observe that our interviewees did not finish their courses on modelling with the perception that modelling tools are bad, as is commonly reported in literature on modelling education, e.g., in \cite{Liebel2016Impact,akayama2013tool}.
Such a negative perception could lead to a reduced uptake of modelling in industry.

Our interviewees raise several challenges that relate to a lack of guidance, feedback, and clarity when it comes to modelling.
While they primarily perceive this as an issue in how assignments are set up and how the courses are designed, the challenges resonate with those in industry.
For instance, a lack of training and guidance is raised in several empirical studies on modelling in industry, e.g., \cite{gorschek2014use,torchiano13,liebel18survey}.
Whittle et al.~\cite{Whittle2013Industrials} highlight that organisational and process factors play a major role in the use of MDE.
Similar issues are raised by our interviewees in the educational context, namely that modelling assignments need suitable processes that allow for iterations and quick feedback loops.
This is especially important as models were not used for automated tasks in any of our cases, i.e., there was no automatic feedback generated.

An important difference of the educational setting to industry is that students often lack both the technical background, e.g., in terms of programming experience, and the domain knowledge required for the example domains.
Therefore, they struggle to contextualise the value and the quality of their models, often leading to a perception of doing a meaningless drawing task.
Publications on modelling education sometimes argue for the use of realistic examples and caution against toy examples, e.g., \cite{kolovos2016towards}.
However, also realistic examples have their pitfalls.
While unrealistic domains are just that, unrealistic, they can also expose students to a known domain, or at least a low level of complexity due to a domain that is artificial and potentially more controlled and restricted.
From this point of view, toy examples can be a suitable tool for modelling education.
Given our findings in this study, we would advise to use primarily example domains the students have sufficient knowledge of.
Whether or not these domains are then simplified artificially or not should depend primarily on context factors in the course, e.g., the time dedicated to understand and work with the domain, the depth of the covered modelling concepts, and the student level.
Another option is to let students choose a domain on their own, which can potentially increase their motivation.
However, they run the risk of selecting a domain which is particularly unsuited given the course assignments.

Finally, we observe that several benefits and challenges voiced by our interviewees are directly reflected in industrial surveys on the use of models, e.g., the usefulness of models for communication \cite{liebel18survey,gorschek2014use,torchiano13}, or the sentiment that models are not worth the effort \cite{gorschek2014use,torchiano13}.
While we are not able to answer this based on our study, we would like to highlight the possibility that these opinions are shaped during university education and then maintained later on.
That is, addressing these issues while educating modellers could lead to a higher uptake and appreciation of modelling in industry.

\section{Conclusion}
\label{sec:Conclusion}
We conducted a case study investigating the perceptions students have of software modelling.
To do so, we conducted 21 interviews with students and instructors in five courses at 3 universities, and consulted assignment descriptions of three of the five courses.
The results offer rich insights into how students perceive modelling.
While related work on the topic exists, it is usually in the form of survey studies, opinion and experience papers, and industrial case studies.

Several findings from related work are confirmed in our study, e.g., that while seeing the benefit of modelling for planning or communication, many students perceive modelling as not worth the effort or as an outright waste of time.
They struggle with the fast pace of courses and a lack of repetition and in-depth feedback that goes beyond the diagram syntax.
However, we additionally find differences to existing literature, or deeper explanations of phenomena observed in related work.
For instance, our interviewees reported no tooling-related challenges, potentially as they were allowed to choose their tool of choice in all cases, and as none of the five courses used any model-based techniques that would require a specific CASE tool.
Finally, we find that students struggle to understand modelling, as they are often at the same time lacking knowledge in areas related to the course projects, i.e., the problem domain, the intended task (such as design or architecture), object orientation, and programming skills.
As there is often no automatic feedback from modelling tools, this means students cannot anchor their models to any known domain.

Our study is of interest to university researchers, educators and professionals.
Specifically for educators, it confirms many existing beliefs, but also highlights new details that can be considered when teaching modelling, i.e., a more nuanced view on how the choice of a problem domain affects the students' perceptions of modelling.
For modelling researchers, it can open up new directions on how to improve guidance and training in modelling, something that also professionals regularly report as challenging.
Finally, professionals can use our findings to better understand what experiences university graduates have with modelling, which benefits they perceive, and why this is the case.

\section*{Acknowledgement}
We would like to thank the interviewees for participating in the study.

\section*{Declarations}
\subsection*{Competing Interests}
The authors declare that they have no conflict of interest.

\subsection*{Data Availability}
We publish 13 out of the 21 anonymised interview transcripts on Zenodo, \url{https://doi.org/10.5281/zenodo.6913780}. We did not receive consent to publish the remaining 8 transcripts.

\bibliographystyle{spmpsci}      
\bibliography{ref}

\appendix
\section{Interview Guides}
\label{app:Interview_Guide}
We present two interview guides in this section used for students and instructors. We scheduled 40 minutes for student interviews and 30 minutes for instructor interviews. Interviews were recorded with permission from the interviewees. 
\subsection{Student Interview Guide}
The interview starts with a small introduction of 5 minutes presented by the interviewer which consists the following points:
\begin{itemize}
    \item Asking permission to record the interview.
    \item Introduction of the interviewer.
    \item Explain the purpose of the study, the research questions. 
    \item Explain the rules of the interview. 
\end{itemize}
Following the introduction, the interview starts. It has two parts, first the \emph{Introductory Questions}. The duration of this part is 10 minutes.
\begin{enumerate}
    \item I1: Which degree are you currently pursuing? (in which major?)
    \item I2: Have you worked in industry? If yes, what role/domain?
    \item I3: Do you have any experience with modelling? What kind of experience do you have with modelling?
    \item I4: Go in detail with the experience, ask about syntax, diagrams that they had used before.
\end{enumerate}

In the second part, \emph{Modelling Questions} we go in detail with the modelling challenges and the duration is a maximum of 25 minutes. 
Before going into the modelling questions, the interviewer explains software modelling and her research interest in software modelling. The reason is to make sure that the interviewee has a clear idea and can give concrete answers.  
\begin{enumerate}
    \item M1: What experience do you have in software modelling?
    \item M2: What did you like from your experience?
    \item M3: What challenges did you face during your experience?
    \item M4: In your opinion what are the reasons behind these challenges?
    \item (If needed): When you are using a modelling notation what typically troubles you?     \item M5: How was your experience with modelling tools? Tell me about the tools you have used in the past.
    \item M6: What did you like and dislike about each of those tools?
    \item M7: For which purposes/in which situations (in SE) do you think modelling is useful? When is this not the case?
    \item M8: In your case can you tell me of an experience where you found modelling useful?
    \item M9: What advantages and disadvantages do you see in modelling?
    \item M10: Did you get enough knowledge about modelling from the courses you have taken on the topic?
    \item M11: What was missing from those courses?
    \item M12: How would the perfect modelling course look like?
    \item M13: Do you have any comments or suggestions?
\end{enumerate}
\subsection{Instructor Interview Guide}
\begin{enumerate}
    \item Could you shortly describe which topics you cover in your course?
    \item Regarding modelling, what are you covering in the course?
        \begin{enumerate}
            \item Are you teaching modelling notation (e.g., UML diagrams)?
            \item Are you teaching how to draw those models (example: identify nouns/verbs in a text to draw class diagrams)?
            \item Are you teaching how the models relate to code?
            \item Are you teaching semantics of modelling notations?
        \end{enumerate}
    \item Roughly how much time is spent on modelling in your course?
    \item Regarding modelling, what is covered in assignments or exam?
    \item How do you think students perceive the modelling part of the course?
    \item What do you think are the things that students struggle with (with respect to modelling)?
    \item What challenges are you facing when teaching modelling?
    \item How do you think this could be improved?
\end{enumerate}

\end{document}